\documentstyle[11pt,twoside,newpasp,epsf]{article}

\begin{document}

\title{IDA: A new software tool for {\it INTEGRAL} field spectroscopy {\it Data Analysis}}

\author{B. Garc\'{\i}a-Lorenzo\altaffilmark{1,2}, J.A. Acosta-Pulido\altaffilmark{2}, and E. Megias-Fern\'andez \altaffilmark{2,3} }

\affil{Isaac Newton Group of Telescopes, 38700-S/C de La Palma, Spain}
\affil{Instituto de Astrof\'{\i}sica de Canarias, 38200-La Laguna, Tenerife, Spain}
\affil{Instituto de Astrof\'{\i}sica de Andalucia, Granada, Spain}

\begin{abstract}

We present a software package, IDA, which can easily handle two-dimensional spectroscopy data. IDA has been written in IDL and offers a window-based interface. The available tools can visualize a recovered image from spectra at any desired wavelength interval, obtain velocity fields, velocity dispersion distributions, etc.

\end{abstract}

\section{Introduction}

The evaluation of 3D data obtained by two-dimensional spectroscopy is not a
 very simple task due to the huge amount of data that they contain. Data analisys is one of the major problems we may have when working with two-dimensional spectroscopy.

IDA is a software tool designed to analysis INTEGRAL data, although can be used with other similar data.

{\bf The instrument:}INTEGRAL (Arribas et al. 1998) is a fiber optical system at the 4.2m William Herschel Telescope (WHT) at the Observatorio del Roque de los Muchachos (ORM), La Palma, Spain. INTEGRAL currently has three different fiber bundles (STD1, STD2, and STD3), which allow different spatial coverage and resolutions. At the focal plane of the telescope, the fibres of each bundle are arranged in two groups, one forming a rectangle, and the other an outer ring for collecting background light. When an extended object is observed with this system, each fiber of the central array receives light from the object, while the ring fibers collect background light (see Arribas et al. 1998 for an extensive description of the instrument). 

The three INTEGRAL bundles can be interchanged on-line depending on the scientific program or the prevailing seeing conditions. IDA automatically recognises from which INTEGRAL bundle the data come from. Although IDA was created for manipulating INTEGRAL data, it can be used with other similar data simply adding an ASCII file that we will explain in section 3.1.

\begin{figure}
\plotfiddle{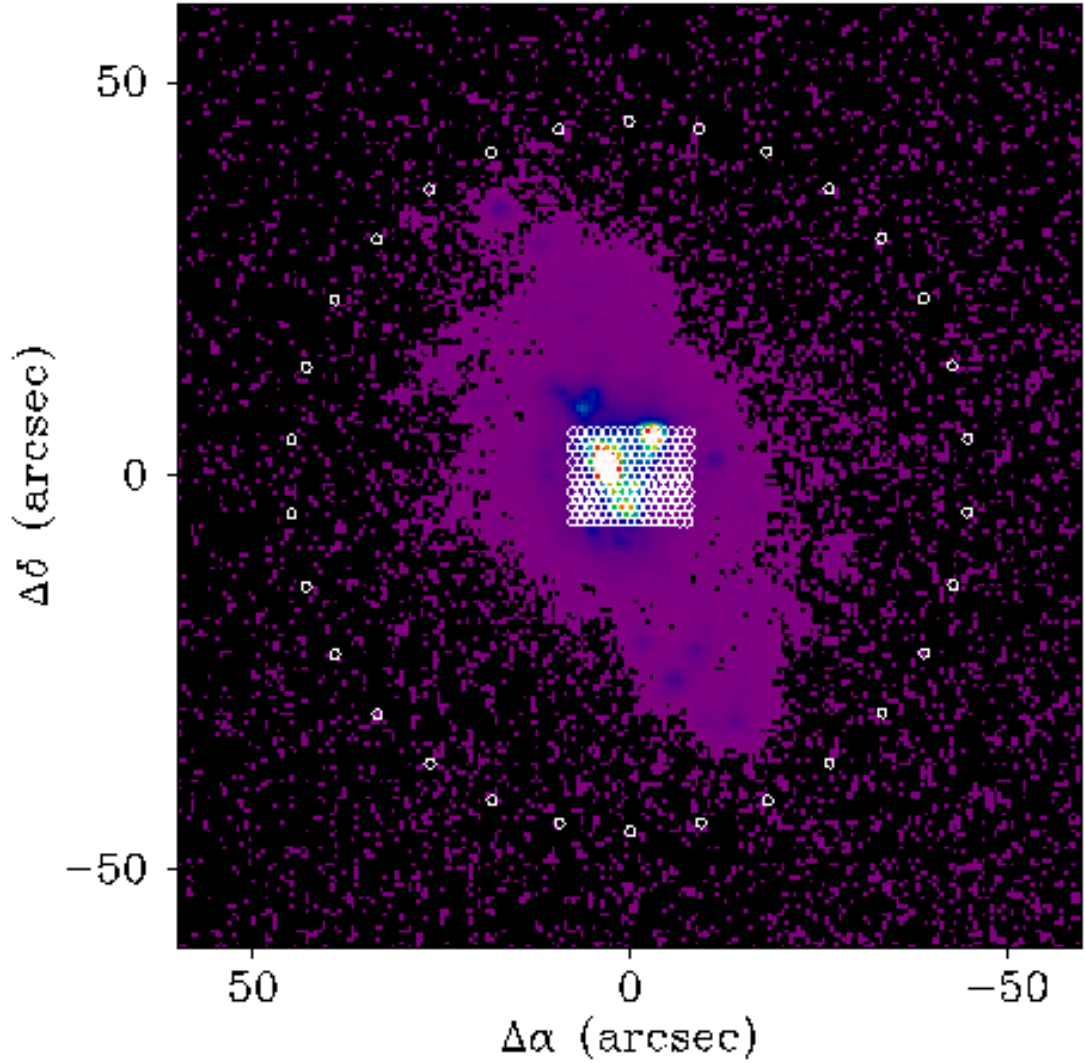}{15cm}{0}{90}{90}{-300}{-140}
\caption{MRK 370 H$\alpha$ filter image taken with the 2.2m telescope at CAHA. The spatial distribution of the STD2 fibers of INTEGRAL at the focal plane has been overlaid. The other two INTEGRAL bundles show a similar distribution at the focal plane.}
\label{integral}
\end{figure}

\section{IDA functions}

IDA is written in IDL {\footnote{IDL is a registered trademark of Research Systems, Inc.}} and needs the IDL environtment to be installed. However, it is not necessary for the user to have any knowlegde of this lenguage. IDA is designed to be interactive and offers a widget base interface.

\begin{figure}
\plotfiddle{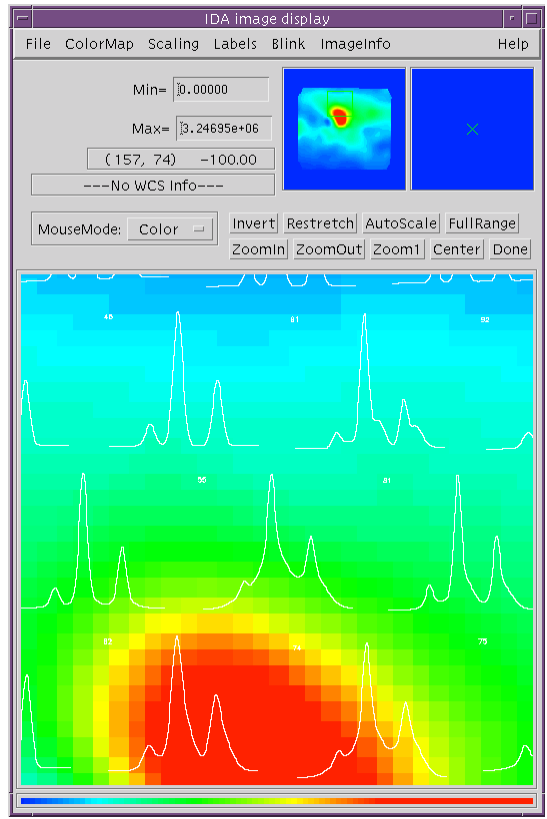}{10cm}{0}{40}{40}{-130}{-10}
\caption{The IDA graphical tool. A NGC 4388 recovered continuum maps is plotted. The observed spectra in the H$\alpha$+[NII] have been overplotted and zoom.}
\label{atv}
\end{figure}

IDA uses an interactive display tool written in IDL (Barth 2001). It allows control of images similar to other display tools like ximtool or ds9 in IRAF (Tody 1986). More information about this graphical tool can be found in the web page http://www.harvard.edu/$\sim$abarth/atv/atv.html. Figure \ref{atv} shows how the display tool looks like when you are using it within IDA.

\begin{figure}
\plotfiddle{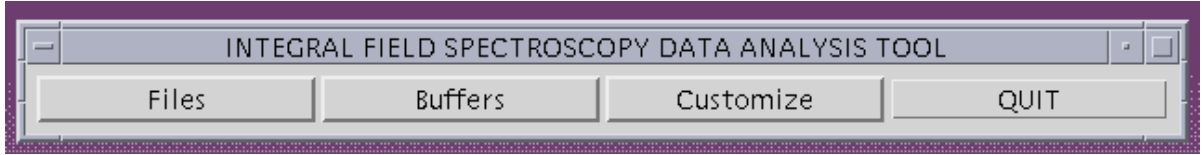}{2cm}{270}{70}{70}{-280}{250}
\caption{The IDA main menu. With this three basic options it is possible to analyse 3D data and also coustomize the setting up for spectral analysis.}
\label{main}
\end{figure}

After starting up IDA, the program offers the user three basic options within the main menu (figure \ref{main}): (1) The {\bf file} option is for reading information into IDA memory; (2) the {\bf buffer} option gives information about data in memory and allows the data analysis itself; and (3) the {\bf customise} option is to modify information and customise your own setting up for spectral analysis.

\subsection{Data reading formats}
\begin{figure}
\plotfiddle{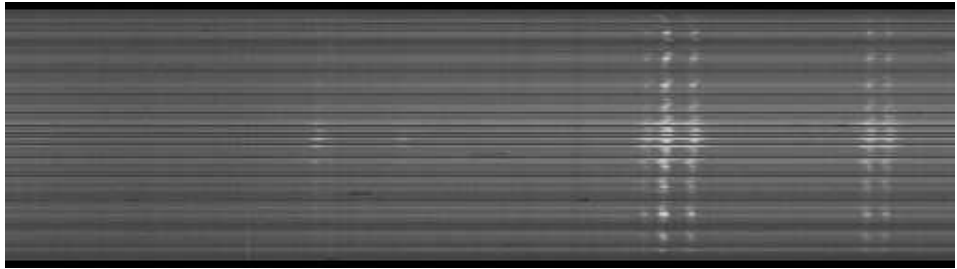}{4cm}{270}{70}{70}{-270}{260}
\caption{Integral reduced data frame. Spectral direction is along X-axis
whereas spatial direction (fibers) is along Y-axis.}
\label{data}
\end{figure}

The {\bf file} option allows reading three different kind of data format:
\begin{itemize}
 \item The first one is a multi-spectrum image that can be a fits or an IRAF image. This image is or should be an INTEGRAL reduced data frame (figure \ref{data}). In the X direction, these images have the spectral information, while each pixel in the Y direction contain the spectrum from an specific fiber. To completely understand these data, IDA needs an ASCII file with the actual position of each fiber in the bundle array. This ASCII file has four columns. The first column indicates the fiber number or the Y pixel in the multi-spectrum frame. The second and third columns are the relative coordinates in arcsec for each fiber; and the fourth colum indicates if the fiber is broken or is available to be considerer for the data analysis.

IDA automatically recognises if the multi-spectrum images come from INTEGRAL, having in memory the corresponding ASCII file. If data are not from INTEGRAL,  IDA will ask for an ASCII file to complete the spatial information of the input data.

\item The second format that can be read by IDA is ASCII. We have included three options to read ASCII according to the origin of these files, due to historical reason of INTEGRAL users.

\item The last option is for reading images created in a previous IDA session, e.g. continuum or intensity maps, velocity fields, etc.
\end{itemize}

\begin{figure}[ht]
\plotfiddle{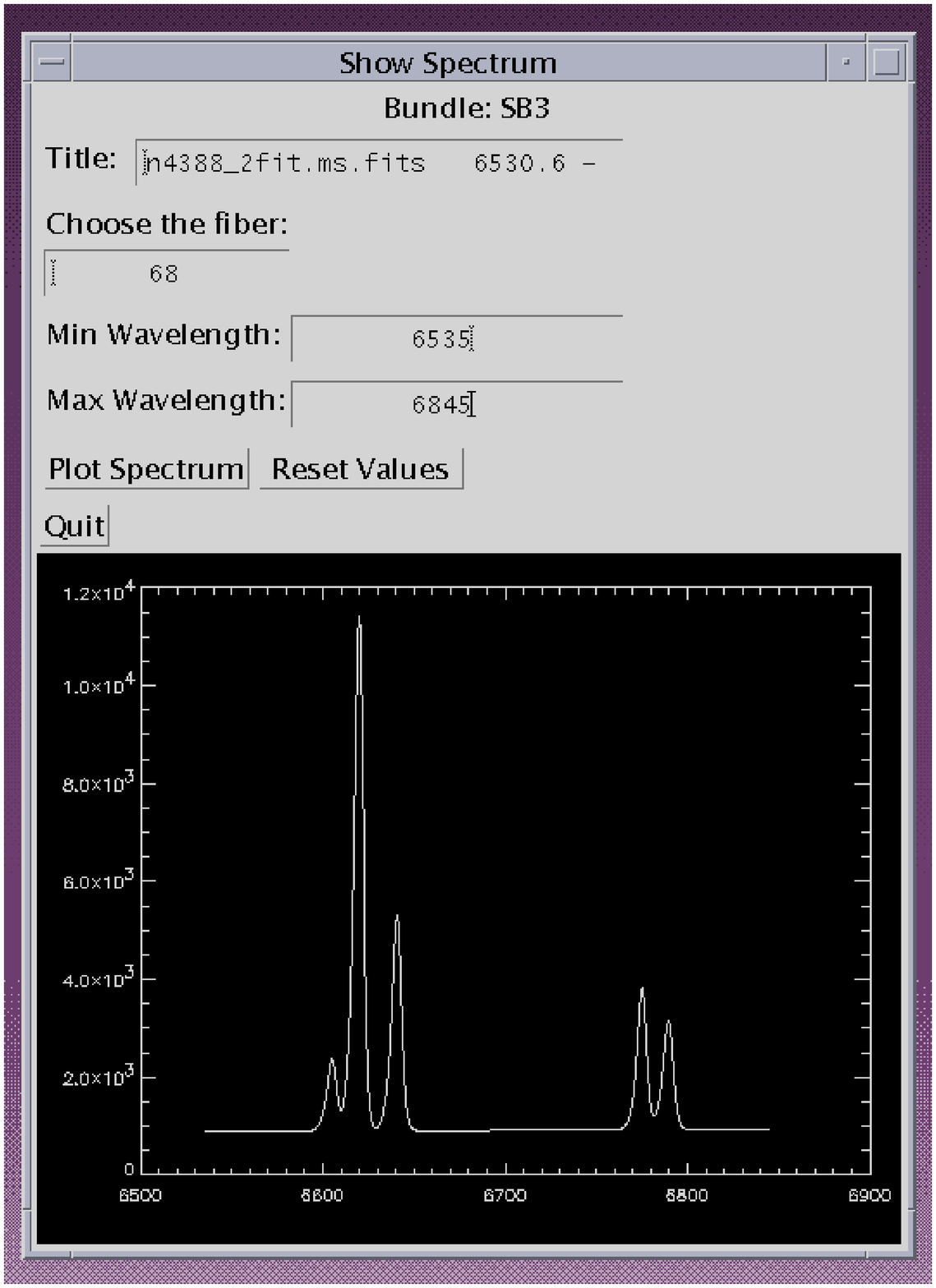}{12cm}{0}{40}{40}{-230}{-30}
\plotfiddle{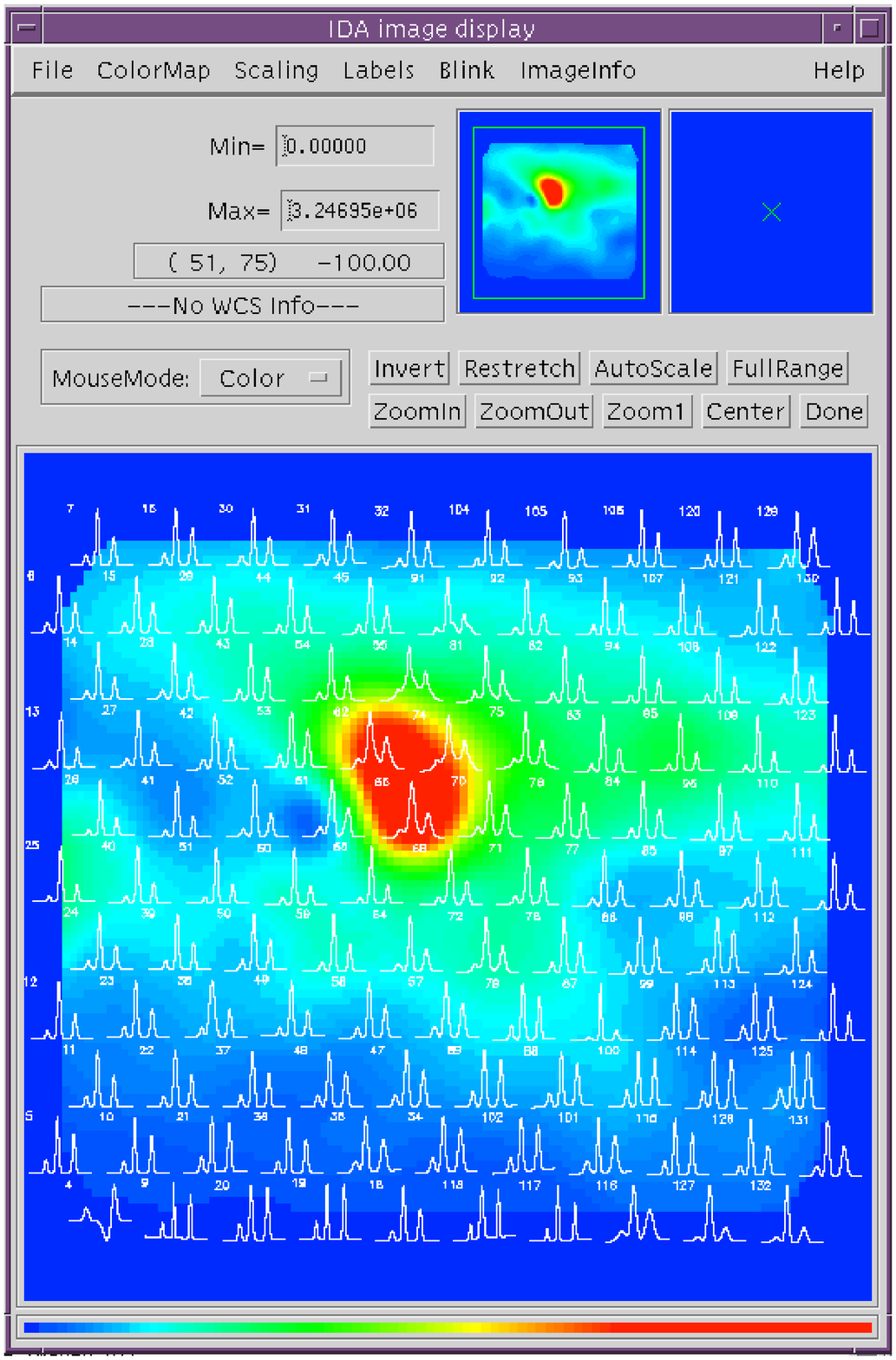}{0cm}{0}{40}{40}{0}{0}
\caption{(left) IDA example of show spectrum; (right) The IDA graphical tool where a NGC 4388 recovered continuum maps is plotted. The observed spectra in the H$\alpha$+[NII] are overplotted.}
\label{fiber_spectrum}
\end{figure}

\subsection{IDA functionality}
\begin{figure}
\plotfiddle{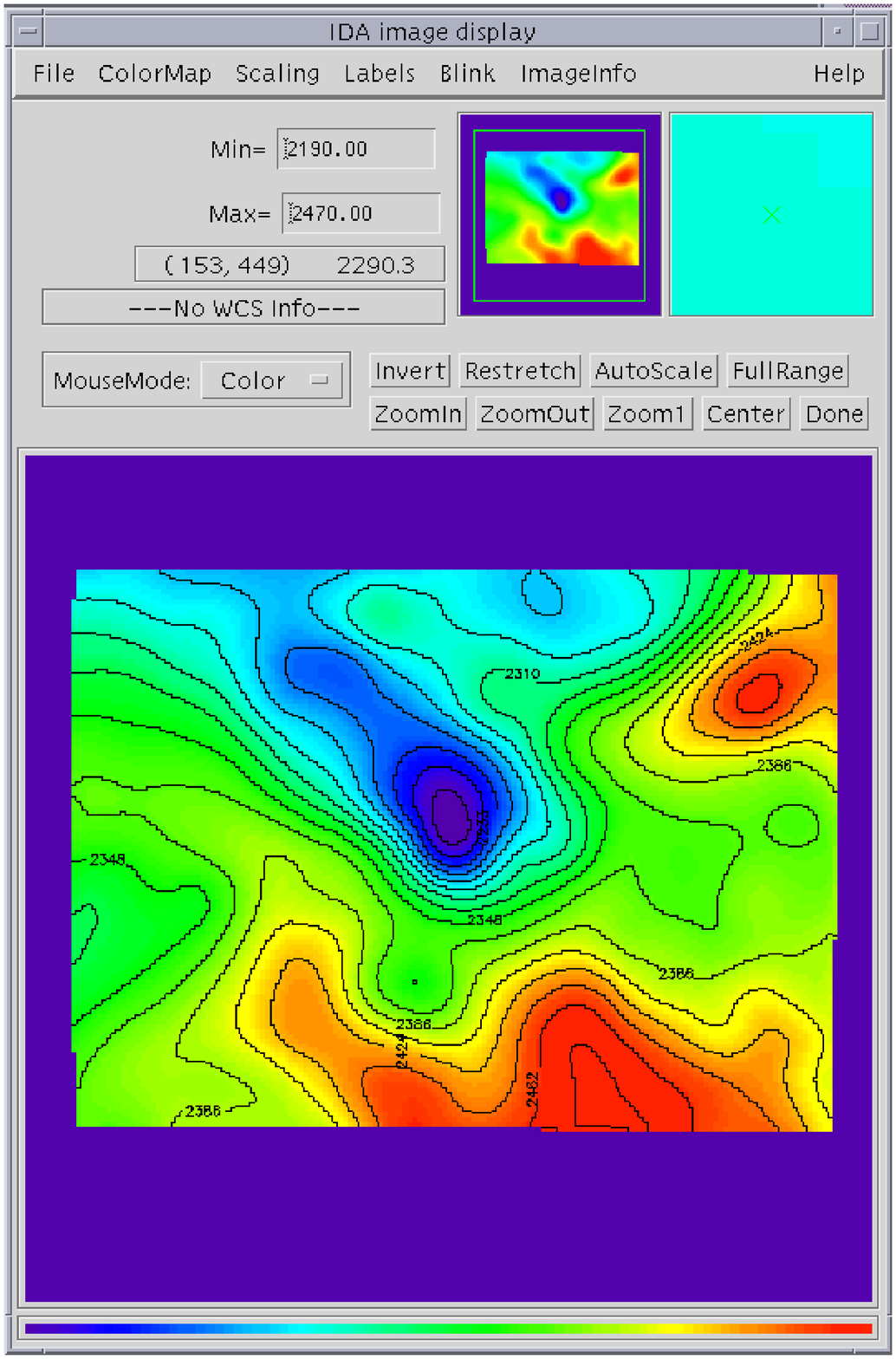}{12cm}{0}{40}{40}{-230}{-30}
\plotfiddle{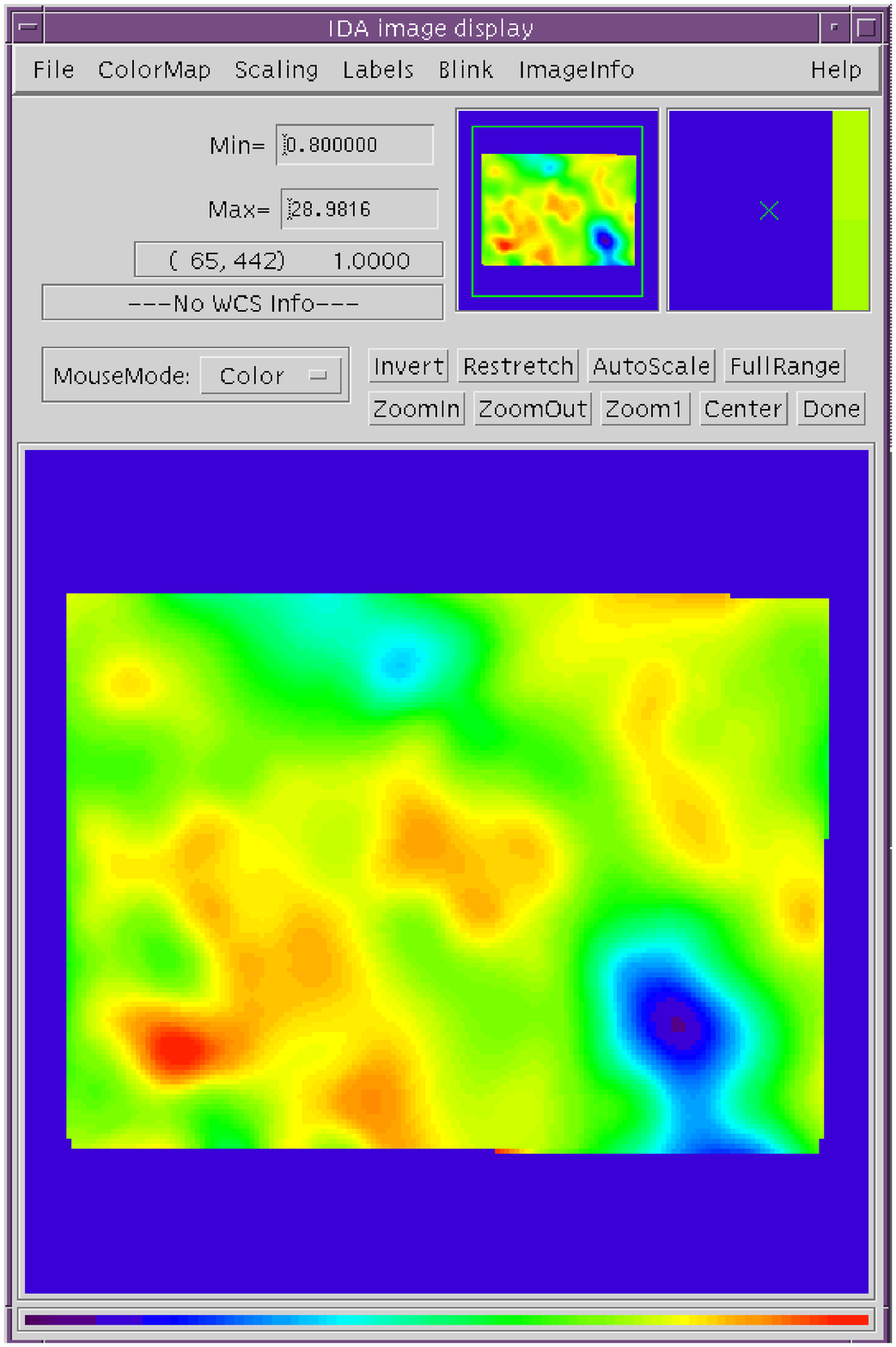}{0cm}{0}{40}{40}{0}{-5}
\caption{(left) NGC2992 ionised gas ([OIII]) velocity field obtained using IDA. (right) The [OIII]/H$\beta$ 2D ratio. This 2D distribution of ionisation has been obtained dividing the [OIII] and H$\beta$ recovered images. These maps can be compared with the ones obtained with starlink routines (Garc\'{\i}a-Lorenzo, Arribas, \& Mediavilla 2001). }
\label{velocidad}
\end{figure}

After reading data, we have several option within IDA. It could be interest to show the spectrum from a specific fiber (figure \ref{fiber_spectrum}), e.g. to define the continuum spectral ranges or to identify lines. But if we are interested for an inspection of all the spectra, we can plot all the spectra together (figure \ref{fiber_spectrum}).

It is also possible to recover maps from spectra. By using a two-dimensional interpolation method to transform the ASCII file with the actual position of the fibers and the value of any spectral feature for each fiber into a regularly spaced rectangular grid. In this way, we can create continuum maps integrating the signal in an specific range,  interactively introducing the pixel scale of the recovered image. Over any recovered image, we can also plot the spectra, e.g. to see if any morphological feature in the map is related to any asymmetry in the profiles. We can also zoom these plots to better see the profile shapes and how they change from one position to another (see figure \ref{atv}).

We can obtain velocity fields (figure \ref{velocidad}) and velocity dispertion distributions using the cross-correlation method. As template we can select the spectrum of a fiber, or any other template from the disk. We can plot the cross-correlation function for any fiber, and we can save the results in an ASCII file. When the cross-corralation is finished, IDA automatically create the velocity field and the velocity dispertion distribution maps.

IDA also works with recovered images. For example, we can divide two intensity maps to obtain ionisation maps (figure \ref{velocidad}) or extiction maps, etc.

IDA can save in disk any recovered image, with an specific header, and can produce camera-ready postcript files for your posters or papers. Cair\'os et al. and Garc\'{\i}a-Lorenzo et al. (these proceedings) present results obtained using IDA for the analysis of their INTEGRAL data.

We are still improving IDA, but it will be available for everybody very soon.

\acknowledgments
We appreciate Antonio Eff-Darwich's helps with the manuscript. The WHT is operated on the island of La Palma by ING in the Spanish Observatorio del Roque de Los Muchachos of the Instituto de Astrof{'{\i}sica de Canarias.

\end{document}